\documentclass[aps,pra,twocolumn,superscriptaddress]{revtex4}

\usepackage{amsmath,amsfonts,graphicx}

\newcommand{\Tr}[1]{\mathrm{Tr}\bigl\{#1\bigr\}}
\newcommand{\tr}[1]{\mathrm{tr}\bigl\{#1\bigr\}}
\newcommand{\avD}{\overline{D^2}}
\newcommand{\ket}[1]{|#1\rangle}
\newcommand{\bra}[1]{\langle #1|}
\newcommand{\braket}[2]{\langle #1|#2 \rangle}
\begin{document}

\title{Mutually unbiased bases in dimension six: The four most distant bases}
\author{Philippe \surname{Raynal}}
\affiliation{Centre for Quantum Technologies, %
National University of Singapore, 3 Science Drive 2, 117543, Singapore}
\author{Xin \surname{L\"u}}
\affiliation{Centre for Quantum Technologies, %
National University of Singapore, 3 Science Drive 2, 117543, Singapore}
\affiliation{Department of Physics, %
National University of Singapore, 2 Science Drive 3, 117542, Singapore}
\author{Berthold-Georg \surname{Englert}}
\affiliation{Centre for Quantum Technologies, %
National University of Singapore, 3 Science Drive 2, 117543, Singapore}
\affiliation{Department of Physics, %
National University of Singapore, 2 Science Drive 3, 117542, Singapore}

\date{5 March 2011}

\begin{abstract}
We consider the average distance between four bases in dimension six. 
The distance between two orthonormal bases vanishes when the bases are the
same, and the distance reaches its maximal value of unity when the bases are
unbiased. 
We perform a numerical search for the maximum average distance 
and find it to be strictly smaller than unity. 
This is strong evidence that no four mutually unbiased bases exist in
dimension six.  
We also provide a two-parameter family of three bases which, 
together with the canonical basis, reach the numerically-found maximum of the 
average distance, and we conduct a detailed study of the structure of
the extremal set of bases.
\end{abstract}

\maketitle

\section{Introduction}
Two orthonormal bases of a Hilbert space are said to be unbiased if the
transition probability from any state of the first basis to any state of the
second basis is independent of the two chosen states. 
In the finite dimensional case of $\mathbb{C}^d$, the normalization of the two
basis states $|a_i\rangle$ and $|b_j\rangle$ of two unbiased bases implies the
defining property 
\begin{eqnarray}
|\langle a_i|b_j \rangle|^2=\frac{1}{d}\quad\mbox{for all $i,j=1,2,\dots,d$}\,.
\end{eqnarray}
This maximum degree of incompatibility between two bases 
\cite{weyl27,schwinger60}
states that the corresponding nondegenerate observables are complementary.
Indeed, the technical formulation of Bohr's Principle of Complementarity
\cite{bohr27} that is given in Ref.~\cite{sew91} relies on the unbiasedness of
the pair of bases.
Textbook discussions of this matter can be found in
Refs.~\cite{schwinger01,englert06}, and Ref.~\cite{debz10} is a recent review
on mutually unbiased bases (MUB), which are sets of bases that are pairwise
unbiased. 

In addition to playing a central role in quantum kinematics, we note that MUB
are important for quantum state tomography \cite{ivanovic81,wootters89}, 
for quantifying wave-particle duality in multi-path
interferometers~\cite{ekkc08}, and for various tasks in the area of quantum
information, such as quantum key distribution \cite{gisin02} or quantum
teleportation and dense coding \cite{durt04b,klimov09,revzen09}.   
In the context of quantum state tomography, 
$d+1$ von Neumann measurements provide
${d-1}$ independent data each in the form of $d$ probabilities with unit sum, 
so that in total one has the required ${d^2-1}$ real numbers that characterize
the quantum state.
A set of ${d+1}$ MUB is optimal, in a certain sense \cite{wootters89}, for
these measurements---if there is such a set.
Such a set is termed \emph{maximal}; there cannot be more than ${d+1}$ MUB,
since there are at most ${d+1}$ ${(d-1)}$-dimensional orthogonal subspaces in
a ${(d^2-1)}$-dimensional real vector space \cite{wootters89}. 

Ivanovic~\cite{ivanovic81} gave a first construction of maximal sets of MUB if
the dimension $d$ is a prime, and Wootters and Fields~\cite{wootters89} 
succeeded in constructing maximal sets when $d$ is the power of a prime.
These two cases have been rederived in various ways; see
Refs.~\cite{bandyopadhyay02,klappenecker04,durt04a}, for example. 
For other finite values of $d$, maximal sets of MUB are unknown, but it is
always possible to have at least three MUB (see \cite{debz10} and references
therein).

The smallest non-prime-power dimension is ${d=6}$.
Little is known for sure about the six-dimensional case, for which Zauner has
conjectured that no more than three MUB exist \cite{zauner99}. 
Numerical studies seem to support Zauner's conjecture
\cite{grassl04,weigert08}. 
Computer-aided analytical methods, like Gr\"obner bases or SemiDefinite
Programming, have also been applied to this problem \cite{weigert10}, 
but limitations in
computational power have so far prevented any definitive answer. 

Recently, Bengtsson \textit{et al.\/} \cite{bengtsson06} introduced  a
distance between two bases for a quantification of the notion of 
``unbiasedness.''
The distance vanishes when the two bases are identical and attains its maximal
value of unity when they are unbiased. 
One can then consider the average squared distance (ASD) between several
bases and search for its maximal value.  
Importantly, this ASD is unity if the bases are pairwise unbiased, and only
then.  
A numerical search for the maximum of the ASD between
four bases in dimension six can be performed. 
Actually, a numerical study on essentially the same quantity was recently
carried out by Butterley and Hall \cite{butterley07}. 
In terms of the ASD, they found the surprisingly large but
strictly-less-than-one maximal value of $0.9983$. 
This is strong evidence that no more than three MUB exist in dimension six. 
However, the set of bases behind this maximum value is not reported in
Ref.~\cite{butterley07}.

It is the objective of the present paper to close this gap. 
In Sec.~\ref{section2} we review the notion of Bengtsson \textit{et al.\/} for
the distance between bases. 
We perform a numerical search for the maximum ASD
between four bases in dimension six and report, in Sec.~\ref{section3},
our results which confirm the maximum found by Butterley and Hall. 
We then provide a two-parameter family of three bases which, together with the
canonical basis, reaches the numerically-found maximum,
for which we give a closed expression. 
We study this family in detail in Sec.~\ref{section4} and 
conclude with a summary and outlook.
Some matters of a technical nature are reported in the Appendix.

\section{A distance between bases}\label{section2}

The main goal of this paper is twofold. First we numerically search for the
maximum value of the ASD between four bases in dimension six and see that we
cannot obtain four MUB. 
And second, we provide a two-parameter family of three
bases which, together with the canonical basis, reaches the numerically-found
maximum. 

Following Bengtsson \textit{et al.\/} \cite{bengtsson06}, 
we consider two orthonormal bases of
kets of $\mathbb{C}^d$,  $a=\{|a_i\rangle\}$ and $b=\{|b_j\rangle\}$, and
quantify their squared distance by
\begin{eqnarray}\label{distance}
D^2_{ab}
&=&1-\frac{1}{d-1}\sum_{i,j=1}^d\biggl(|\langle a_i|b_j \rangle|^2
                                      -\frac{1}{d}\biggr)^2
\nonumber\\ 
&=&\frac{1}{d-1}\sum_{i,j=1}^d|\langle a_i|b_j \rangle|^2
                            \Bigl(1-|\langle a_i|b_j \rangle|^2\Bigr)\,.
\end{eqnarray}
Clearly, this distance is symmetrical,
$D_{ab}=D_{ba}$ and vanishes when the bases are the same, that
is: when the two sets of projectors $\{\ket{a_i}\bra{a_i}\}$ and
$\{\ket{b_j}\bra{b_j}\}$ are identical; the maximal distance is unity,
$D_{ab}\leq1$; and this maximum is reached if the bases are unbiased,
$|\langle a_i|b_j \rangle|^2=1/d$, and only then.

In the original reasoning by Bengtsson \textit{et al.\/}, 
$D_{ab}$ is actually the
chordal Grassmanian distance of two planes in the ${(d^2-1)}$-dimensional real
vector space associated with traceless hermitian operators in the
$d$-dimensional complex Hilbert space.
One can also view $D_{ab}$ as the Hilbert-Schmidt distance between two
rank-$d$ statistical operators in $\mathbb{C}^d\otimes\mathbb{C}^d$ that are
in one-to-one correspondence with the $d$-dimensional bases.
Consult Appendix~\ref{app1} for this matter. 

For a set of $k$ bases, we have the ASD between the
$k(k-1)/2$ pairs of bases, given by
\begin{eqnarray}
\avD=\frac{2}{k(k-1)}\sum_{a<b=1}^k D_{ab}^2\,.
\end{eqnarray} 
Owing to the normalization, we have $\avD\leq1$ with $\avD=1$ only if the $k$
bases are pairwise unbiased.

With this notion of distance at hand, we can numerically search for the
maximum ASD between four bases in dimension six and see whether
we obtain ${\avD=1}$, or in other words, if we can find four MUB. 
This search is the subject matter of the next section.

\section{Numerical Study}\label{section3}
Our numerical approach relies on the mapping between 1-qudit operators and
2-qudit states established in Chapter 3 of \cite{debz10}. 
Plus we use the steepest-ascent algorithm to find the maximum ASD
between four bases in dimension six. 
Details of the numerical method are presented in Appendix~\ref{app1}. 
Our numerical results are reported below. 

A similar numerical study was recently performed by Butterley and Hall
\cite{butterley07} who  minimized ${1-\avD}$ with the so-called
Levenberg-Marquadt algorithm.
Our approach confirms the extremal value they found, and we also
exhibit the structure of the four bases that maximize $\avD$ for ${d=6}$. 

We have used our code not only in dimension ${d=6}$ but also for other $d$
values as a mean of benchmarking. 
We have run our code 2,500 times for the dimensions two to five, 10,000 times
for the dimension six and 300 times for the dimension seven, both for
${k=d+1}$ bases and for four bases. 
Our results are summarized in Table~\ref{tab:table1}.

\begin{table}
\caption{\label{tab:table1}%
Rate of success and CPU time (in seconds) for the steepest-ascent search for
the maximum ASD. 
The absolute maximum of ${\avD=1}$ is always reached for 
$d+1$ bases in dimensions $d=2$, $3$, $4$, and $5$. 
As the seven-dimensional case illustrates, the difficulty of finding the global
maximum increases rapidly with the dimension because there are many local
maxima at which the steepest-ascent search can get stuck. 
We have also looked for the largest ASD between 
four bases in dimensions two to seven. 
We could not find four MUB in dimensions two and six.
The CPU time refers to a 
Intel\textsuperscript{\textregistered}Core\texttrademark2~Duo~CPU~E6550 
processor at 2.33\,GHz, supported by 3.25\,GB of RAM.
}
\centering
\begin{ruledtabular}
\begin{tabular}{@{\ }c@{\quad}ccc@{\quad}ccc@{\ }}
 &\multicolumn{3}{l}{\underline{\makebox[95pt][c]{$d+1$ bases}}}
 &\multicolumn{3}{l}{\underline{\makebox[95pt][c]{4 bases}}}\\
&&Success&CPU&&Success&CPU\\[-1ex]
$d$&$\avD_{\mathrm{max}}$&rate (\%)&time&
$\avD_{\mathrm{max}}$&rate (\%)& time\\ \hline
2&1&100&0.049&8/9&100&0.108\\ 
3&1&99.9&0.272&1&99.9&0.272\\ 
4&1&100&1.268&1&100&0.976\\ 
5&1&99.7&4.432&1&59.8&10.995\\ 
6&0.9849&39.2&188.407&0.9983&69.6&20.158\\ 
7&1&3.8&467.157&1&1.1&101.002
\end{tabular} 
\end{ruledtabular}
\end{table}

Only in two cases, the maximum ASD does not reach the upper bound
of ${\avD=1}$. 
They are the cases of four bases in dimension two and six.

At most three MUB can be constructed in dimension two. 
Thus the maximum ASD between four bases has to be less than one. 
This example is interesting because it can be analytically solved. 
In $\mathbb{R}^3$, the four bases correspond to the tetrahedron, 
where each edge represents a basis.

\begin{figure}
\centerline{%
\includegraphics[bb=55 523 294 736]{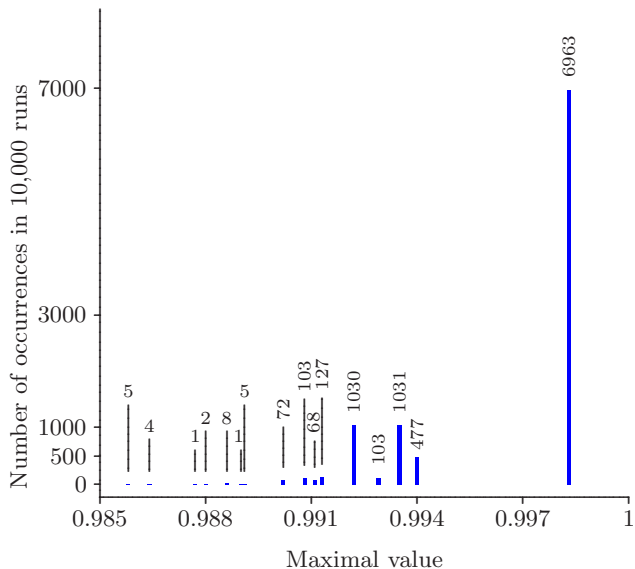}}
\caption{Histogram of the maximum values of the ASD 
found during a numerical search for 10,000 randomly chosen initial four bases.
The search converges to one of the local maxima in about 30\% of all
runs, and to the global maximum of ${\avD_{\mathrm{max}}=0.9983}$ for the
other 70\% of initial bases. 
}
\label{fig:histo}
\end{figure}

Importantly, we have searched for the maximum ASD between four
bases in dimension six. 
We have found the largest value to be $\avD_{\mathrm{max}}=0.9983$. 
In the search for the global maximum, we have also found a few other local
maxima whose frequencies of occurrence are reported in
Figure~\ref{fig:histo}.
These results are consistent with those reported by Butterley and
Hall~\cite{butterley07}. 
We find the same local and global maxima with very similar frequencies. 
This is as expected because we have generated the four random bases from which
the search proceeds in the same way as Butterley and Hall, using the same
dedicated Matlab command.  
The two numerical methods are different, however. 
We use the steepest-ascent algorithm while they employ the  Levenberg-Marquadt
algorithm for a nonlinear least-squares optimization. 

Since we consider four bases, there are six pairs of bases and their
respective distances are not without interest. 
Indeed, it turns out that one basis is unbiased with the three remaining
bases.  
And these three remaining bases are themselves equidistant. 
The immediate implication is that the privileged basis can be chosen to be the
computational basis while the three remaining bases are Hadamard bases, that is:
the unitary matrices composed of the columns that represent the basis kets with
reference to the computational basis are complex Hadamard matrices divided by 
$\sqrt{6}$. 
We recall here that a complex Hadamard matrix is a $d$-dimensional square
matrix satisfying the two conditions of unimodularity and 
orthogonality~\cite{hadamard93} 
\begin{eqnarray}\label{Hada}
|H_{ij}|&=&1\quad\mbox{for}\ i,j=1,\dots,d\,,\nonumber\\
HH^\dagger &=& d \openone\,.
\end{eqnarray}
Therefore, the unitary matrix $H/\sqrt{d}$ has matrix elements that can be 
related to a pair of unbiased bases: ${\braket{a_i}{b_j}=H_{ij}/\sqrt{d}}$.

In addition to maximizing $\avD$, our code also returns the four bases
for which the maximum is achieved. 
After a bit of polishing---the set of four bases is not unique, since global
unitary transformations yield equivalent sets, and the order of kets in each
basis is arbitrary---this allows us to seek for the structure hidden
behind the maximum ASD. 
In the next section we will present a two-parameter family of three bases. 
The two parameters are two phases while the three bases are three Hadamard
bases. 
We study in detail the properties of this family and show that, for some
definite values of the two parameters, these three bases together with the
canonical basis reach the numerically-found maximum ASD of
$0.9983$. 
This definite structure of the optimal four bases is our main result, with
a closed expression for $\avD_{\mathrm{max}}$ as a most-welcome bonus; see
Eq.~(\ref{avDmax}) below. 

Harking back to Table~\ref{tab:table1}, we note that the best set of seven
bases in dimension six has an ASD of $0.9849$, short of 
unity by a mere one-and-a-half percent.
For all practical purposes---those of state tomography, say---these seven
bases are marginally worse than the imaginary seven MUB that no one has
managed to find.

\section{The two-parameter family}\label{section4}

Following Karlsson~\cite{karlsson10}, we express the two-parameter family in
terms of $2\times2$ block matrices where each of the nine blocks is itself a
complex Hadamard matrix. 
Such $2\times2$ block matrices are called $H_2$-reducible. 
The two-parameter family contains three bases, the fourth basis being the
canonical basis. 
We will see that these three Hadamard bases are equidistant, 
that their determinants are identical, and that they belong to the so-called
Fourier transposed family $F_6^T$. 
Finally, we will show that together with the canonical basis they reach the
numerically-found maximum of the ASD.

\subsection{Parametrization}
We begin by defining a few quantities. 
We will need the third root of unity $\omega=\exp(i\,2\pi/3)$ as well as
the following $2\times2$ matrices: 
\begin{equation}
Z=\begin{bmatrix} 1&0\\ 0&-1 \end{bmatrix}\,,
\quad
X=\begin{bmatrix} x^*&0\\ 0&x \end{bmatrix}\,,
\quad
F_2=\begin{bmatrix} 1&1\\ 1&-1 \end{bmatrix} \nonumber
\end{equation}
and
\begin{equation}\label{2x2a}
T=\begin{bmatrix} 1&\omega t^2\\ 1&-\omega t^2\end{bmatrix}
\end{equation}
where $t=\exp(i \theta_t)$ and $x=\exp(i \theta_x)$ are two phases. 
Let us notice that $T$ and $F_2$ are themselves Hadamard matrices.

The Hadamard matrices for the three bases are given by
\begin{eqnarray}
M_1&=&\begin{bmatrix} X&0&0\\ 0& i\omega^* t Z X^{*2}&0\\ 0&0&X \end{bmatrix}
\frac{1}{\sqrt{6}}
\begin{bmatrix}	F_2&F_2&F_2\\ F_2&\omega F_2&\omega^* F_2\\
                T&\omega^* T&\omega T \end{bmatrix} \nonumber \\
&=& \frac{1}{\sqrt{6}} X_1 N_1\,, \nonumber \\
M_2&=& \frac{1}{\sqrt{6}}\begin{bmatrix} F_2&F_2&F_2\\
                                         T&\omega T&\omega^* T\\
                                         T&\omega^* T&\omega T \end{bmatrix}
=\frac{1}{\sqrt{6}} N_2\,, \nonumber
\end{eqnarray}
and
\begin{eqnarray} \label{M1}
M_3 &=& \begin{bmatrix}X^*&0&0\\ 0& \omega^* X^*&0\\ 0&0&-i t ZX^2 \end{bmatrix}
\frac{1}{\sqrt{6}}
\begin{bmatrix}	F_2&F_2&F_2\\ T&\omega T&\omega^* T\\
                F_2&\omega^* F_2&\omega F_2\end{bmatrix} \nonumber \\
&=&  \frac{1}{\sqrt{6}} X_3 N_3\,.
\end{eqnarray}
In the above parameterization, we have introduced the matrices $X_i$ and $N_i$,
$i=1,2,3$, which we will address as dephasing and central matrices,
respectively. 
The derivation of this parameterization is explained in Appendix~\ref{app2}. 

The next section is devoted to proving the three properties earlier
mentioned. 
Before we turn to the proofs, we wish to point out that an additional relation
between the two phases $x$ and $t$ exists,
\begin{equation}\label{fame}
\cos\bigl(\theta_t+\tfrac{1}{3}\pi\bigr)
=\frac{\cos(2 \theta_x)}{\sin(\theta_x)}\,.
\end{equation}
It reduces the two-parameter family to a single-parameter family. 
Of course, as a subfamily, it conserves all the fundamental properties of the
two-parameter family. 
Furthermore it still reaches the maximum ASD.

\subsection{Properties}
\subsubsection{Equidistance}
A significant property of the three proposed Hadamard matrices is their
equidistance. 
The relevant terms that appear in the distance between the two bases $M_a$ and
$M_b$ (i.e., $|\langle a_i|b_j\rangle|$) are the elements of the product
matrix $M_a^\dagger M_b^{\ }$ (i.e., $\langle a_i|b_j\rangle$) in absolute
value. 
Therefore, if the three product matrices $M_1^\dagger M_2^{\ }$, 
$M_2^\dagger M_3^{\ }$ and $M_3^\dagger M_1^{\ }$ have equal coefficients in absolute
value, then the three bases $M_1$, $M_2$ and $M_3$ are equidistant. 
This is exactly what happens here. 
Indeed, we have the following cyclic structure:
\begin{equation}
M_1^\dagger M_2^{\ } =\frac{1}{6}
\begin{bmatrix}
  a_1&a_2&a_3\\
  a_3&a_1&a_2\\
  a_2&a_3&a_1
\end{bmatrix}\,,
\quad
M_2^\dagger M_3^{\ } =\frac{1}{6}
\begin{bmatrix}
  b_1&b_2&b_3\\
  b_3&b_1&b_2\\
  b_2&b_3&b_1
\end{bmatrix}\,, \nonumber
\end{equation}
and
\begin{equation} \label{prod}
M_3^\dagger M_1^{\ } =\frac{1}{6}
\begin{bmatrix}
  c_1&c_2&c_3\\
  c_3&c_1&c_2\\
  c_2&c_3&c_1
\end{bmatrix}
\end{equation}
where, on the one hand, the $2\times2$ submatrices $a_1$, $b_2$ and $c_3$ have
the same coefficients in absolute value and, on the other hand, $a_2$, $a_3$,
$b_1$, $b_3$, $c_1$ and $c_2$ have the same coefficients in absolute value. 
More precisely, these matrices have the following forms 
(where the symbol $\check{\ }$ stands for swapping the two diagonal elements).
First,
\begin{equation}
a_1 =
\begin{bmatrix}
   \alpha & \beta\\
  -\beta^* & \alpha^*
\end{bmatrix}\,,
\quad
b_2 = \check{a}_1\,, \nonumber
\end{equation}
and
\begin{equation}
c_3 =
\begin{bmatrix}\label{coef1}
  i \omega t^* \beta & i \omega^* t \alpha\\
  i \omega t^* \alpha^* & i \omega t^* \beta
\end{bmatrix}\,.
\end{equation}
Second,
\begin{equation}\label{coef2}
  \begin{array}[b]{rclcrcl}
a_2 &=&
\begin{bmatrix}
  \gamma & \delta\\
  \epsilon & \omega^* \gamma^*
\end{bmatrix}\,,
&&
b_1 &=&\check{a}_2\,,\\[3ex] 
a_3 &=&
\begin{bmatrix}
  \omega \gamma & -\epsilon^*\\
  -\delta^* & \gamma^*\end{bmatrix}\,,
&&
b_3& =&\check{a}_3\,, \\[3ex]
c_1 &=&
\begin{bmatrix}
  i t^* \delta & i \omega t \gamma\\
  i \omega^* t^* \gamma^* & -i \omega^* t^* \epsilon^*
\end{bmatrix}\,,
&\quad&
c_2 &=&\check{c}_1\,.
\end{array}
\end{equation}
The various coefficients in Eqs.~(\ref{coef1}) and (\ref{coef2}) can be
expressed in terms of the two angles $\theta_x$ and $\theta_t$, 
\begin{eqnarray}
\alpha &=& 4 \cos(\theta_x) \Bigl(1- \omega t^* \sin(\theta_x)\Bigr)\,,
\nonumber \\
\beta &=& -2 i \omega^* t \Bigl( \cos(2 \theta_x)- 2 \cos(\theta_t - 2\pi/3) 
\sin(\theta_x) \Bigr)\,, \nonumber \\
\gamma &=& -2 \omega^* \cos(\theta_x) 
\Bigl( \omega^* + 2 t^* \sin(\theta_x)\Bigr)\,, \nonumber \\
\delta&=& -2 i t \Bigl( \cos(2 \theta_x)
                       - 2 \cos(\theta_t) \sin(\theta_x) \Bigr)\,, \\
\epsilon &=& -2 i \omega^* t^*  \Bigl( \cos(2 \theta_x)
               - 2 \cos(\theta_t + 2\pi/3) \sin(\theta_x) \Bigr)\,. \nonumber
\end{eqnarray}
When Eq.~(\ref{fame}) is fulfilled, we have $\epsilon=\omega^* \delta^*$ 
and a few simplifications arise. 
We obtain
\begin{equation}
a_3 = \omega Z a_2 Z\,, 
\quad b_3 = \omega Z b_1 Z\,, \quad \textrm{and} \ 
c_2 = - Z c_1 Z\,,
\end{equation}
for example.

\subsubsection{Determinant}
A direct calculation shows that
\begin{equation}
\textrm{Det}(X_1)=\textrm{Det}(N_1)=\textrm{Det}(X_3)
=\textrm{Det}(N_3)=wt^2\,.
\end{equation}
Accordingly, the three Hadamard bases share the same determinant
\begin{eqnarray}
\textrm{Det}(M_1)&=&\textrm{Det}(M_2)=\textrm{Det}(M_3)=w^*t^4\,.
\end{eqnarray}
However, although the determinants are equal, there seems to be no simple
relation between the three matrices $M_1$, $M_2$, and $M_3$. 
In particular, they do not have the same spectrum and are, therefore, not
related by unitary operators.

\subsubsection{Fourier transposed family}
The Fourier transposed family, first studied by Haagerup~\cite{haagerup96}, 
is parameterized by Karlsson in the form~\cite{karlsson10}
\begin{equation}\label{FT6}
F_6^T \sim
\begin{bmatrix}
  F_2&F_2&F_2\\
  T_1&\omega T_1&\omega^* T_1\\
  T_2&\omega^* T_2&\omega T_2
\end{bmatrix}\,,
\end{equation}
where the $2\times2$ Hadamard matrices $T_1$ and $T_2$ are given by
\begin{equation}\label{2x2Zi}
T_i = \begin{bmatrix}  1&t_i\\    1&-t_i \end{bmatrix}\,, \quad |t_i|=1\,.
\end{equation}
The equivalence relation in Eq.~(\ref{FT6}) means equality up to left and right
dephasing and left and right permutations. 
In other words, the central matrix is the fundamental object that specifies the
equivalence class.
In the form of Eq.~\eqref{M1}, it is clear that the three matrices $N_1$,
$N_2$, and $N_3$ belong to the Fourier transposed family. 
As a result, the two-parameter family itself belongs to the Fourier transposed
family. 

Let us note here that only the right equivalence is natural for more than two 
bases as it states that bases are defined up to permutations 
and global phases of their basis states. 
In particular, the distance between bases 
is invariant under right equivalence but not under left equivalence.

\subsection{Average distance}
Let us now compute the global maximum of the ASD between the
three bases. 
Since the three bases are equidistant, we only have to compute the distance
between, say, $M_1$ and $M_2$. 
A direct calculation leads to the following expression
\begin{eqnarray}
D^2_{12}(\theta_x,\theta_t)=\frac{8}{45}
\bigl[5-P\bigl(\sin(\theta_x),\cos(\theta_t +\tfrac{1}{3}\pi)\bigr)\bigr]\,,
\end{eqnarray}
with the polynomial
\begin{eqnarray}
P(p,q) &=& 8 p^8+8 q^2 p^6-16 q^3 p^5\nonumber\\
&&\mbox{}+16 q p^5-16 q^2 p^4+8 q^3 p^3\nonumber \\
&&\mbox{}-7 p^4-14 q p^3+8 q^2p^2\nonumber \\
&&\mbox{}+2 p^2+4 q p\,. 
\end{eqnarray}
We denote by $(p_{\mathrm{opt}},q_{\mathrm{opt}})$ the $(p,q)$ pair for which
$P(p,q)$ is minimal and, therefore, $D_{12}(\theta_x,\theta_t)$ is maximal.
It turns out that $q_{\mathrm{opt}}$ is related to $p_{\mathrm{opt}}$ by
\begin{equation}\label{p->q}
\cos(\theta_t^{\mathrm{opt}} +\tfrac{1}{3}\pi)=q_{\mathrm{opt}}
=\frac{1-2p_{\mathrm{opt}}^2}{p_{\mathrm{opt}}}\,,
\end{equation}
which is a particular evaluation of the function defined in Eq.~(\ref{fame}),
and $p_{\mathrm{opt}}^2$ is the unique real solution of a cubic equation, 
\begin{eqnarray}
112 p_{\mathrm{opt}}^6 - 192 p_{\mathrm{opt}}^4 + 111 p_{\mathrm{opt}}^2 =22\,,
\end{eqnarray}
that is
\begin{eqnarray}
\sin(\theta_{x\vphantom{t}}^{\mathrm{opt}})^2=p^2_{\mathrm{opt}} 
=\frac{3+16r-r^2}{28r}=0.6946
\end{eqnarray}
with $r=(21\sqrt{3}-36)^{1/3}=0.7199$. 
It follows that there are eight optimal pairs of phases
$(\theta_{x\vphantom{t}}^{\mathrm{opt}},\theta_t^{\mathrm{opt}})$ for which
the maximal distance ${D^{\mathrm{max}}_{12}}$ is reached. 
The above expressions for $\theta_x^\textrm{opt}$ and $\theta_t^\textrm{opt}$
can be injected back into the formula of the distance to obtain first
${D^{\textrm{max}}_{12}}$ and then 
\begin{eqnarray}\label{avDmax}
\avD_{\,\mathrm{max}}&=&\frac{1}{70}\bigl[71-12\cos(\theta_x^{\mathrm{opt}})^4\bigr]
\nonumber\\&=&
\frac{1}{70}\biggl[71-3\biggl(\frac{r^2+12r-3}{14r}\biggr)^2\biggr]
=0.9983\,,\quad
\end{eqnarray}
which agrees with the numerically-found maximum ASD within the
machine precision. 

Furthermore, the distance ${D_{12}}$ vanishes for
\begin{eqnarray}
&&\theta_x=\pi/2\,,\quad\phantom{-}\theta_t=0 \ (\mbox{mod}\ 2\pi/3)
\nonumber\\
\textrm{and}&& 
\theta_x=-\pi/2\,,\quad\theta_t=\pi/3  \ (\mbox{mod}\ 2\pi/3)\,.
\end{eqnarray}
As can be verified from the parameterization (\ref{M1}) or from the matrix
products (\ref{prod}), the bases are indeed identical up to global phases and
permutations for these values of the two phases $\theta_x$ and $\theta_t$. 

\begin{figure}
\centerline{\includegraphics{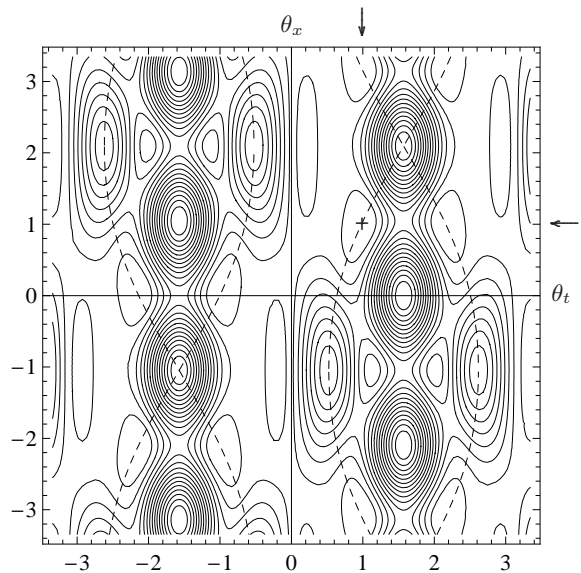}}
\caption{\label{fig:contour}%
Contour plot of the ASD for the two-parameter family.
Along the dashed curves, relation (\ref{fame}) holds.
The four single-parameter families---one for each dashed curve---are
equivalent to the two-parameter family in the sense that the maximal and
minimal value of the ASD can be found by searching along one of
the dashed lines only.
The arrows point to the location of one of the eight maxima at
$(\theta_x,\theta_t)=(0.9852,1.0094)$, marked by a cross.
} 
\end{figure}

We can also consider the single-parameter family that we obtain when
eliminating $\theta_t$ by using Eq.~(\ref{fame}).
Since Eq.~(\ref{p->q}) is equivalent to Eq.~(\ref{fame}), this
single-parameter family reaches the maximum of the ASD --- and also the
minimum since Eq.~(\ref{p->q}) is obeyed by 
$(\theta_x,\theta_t)=(\pi/2,2\pi/3)$.  
This is illustrated in Fig.~\ref{fig:contour}, 
a contour plot of $\avD$ for the two-parameter family of
Hadamard bases, with the location of the $(\theta_x,\theta_t)$ values 
of the single-parameter family indicated. 
The location of one of the eight maxima is marked, and the locations of the
other seven follow from the symmetry properties of the contours.

\section{Summary and outlook}\label{conclusion}
We performed a numerical search for the maximum ASD between
four bases in dimension six. 
We found that it is strictly smaller than
unity and so confirmed the recent study by Butterley and
Hall~\cite{butterley07}.  
We regard this result as strong evidence that no four MUB exist in dimension
six. 

Next, we went beyond this numerical result by providing the four bases
behind the numerically-found maximum. 
More specifically, we found a two-parameter family of three bases, which
together with the canonical basis, reaches the maximum of the ASD.  
We characterized this two-parameter family in full. 
We proved its inclusion in the Fourier transposed family and shown that
the three base are equidistant. 
Furthermore, we analytically computed the maximum ASD
between these three Hadamard bases and the canonical basis to show that it
reproduces the numerical result.

Two directions might be relevant for an extension of the present study. 
First, it would be interesting to see if the optimality of our solution can be
extended to a larger family of bases, for example, to the whole Fourier
transposed family. 
Second and complementarily, there might exist an argument to restrict the
search for the maximum ASD between the canonical basis and three
Hadamard bases to the Fourier transposed family, instead of the entire
Hadamard family which, so far, has not been fully parameterized. 
In this context, however, it should be noted that---as follows from the
findings of Jaming \textit{et al.\/} \cite{Jaming+4:09}---there are no four
MUB if one restricts the search to members of the Fourier family.

Finally, if there is no complete set of seven MUB in dimension six, the
optimal measurement for state tomography, in terms of statistical errors,
remains to be found. 

\begin{acknowledgments}
Centre for Quantum Technologies is a Research Centre of Excellence funded by
Ministry of Education and National Research Foundation of Singapore. 
The authors wish to express their grateful feelings to Ingemar Bengtsson for
enlightening comments and friendly discussions.  
\end{acknowledgments}

\appendix

\section{Numerical method}\label{app1}

As discussed in Sec.~3.1 of Ref.~\cite{debz10}, for any ket $\ket{\varphi}$ or
bra $\bra{\phi}$ in a $d$-dimensional Hilbert space $\mathcal{H}$ or
$\mathcal{H}^{\dagger}$, respectively, there is a
conjugate bra or ket  
\begin{eqnarray} 
\mathcal{H}\ni\ket{\varphi} &\longleftrightarrow& 
\bra{\varphi^{\ast}} \in \mathcal{H}^{\dagger}\,,\nonumber\\
\mathcal{H}^{\dagger}\ni\bra{\phi} &\longleftrightarrow& 
\ket{\phi^{\ast}} \in \mathcal{H} 
\end{eqnarray}
such that
\begin{equation}\label{prescription}
\braket{\varphi^{\ast}}{\phi^{\ast}}=\braket{\varphi}{\phi}^{\ast}
=\braket{\phi}{\varphi}\,.
\end{equation}
This mapping is not unique, but two different realizations differ at most by
a unitary transformation. 
As a rule, $\bra{\phi^*}$ and $\bra{\phi}=\ket{\phi}^{\dagger}$ are different
bras.

Once a particular choice of mapping has been made, there is a one-to-one
correspondence between one-qudit operators and two-qudit kets, 
\begin{equation}{\label{map}}
\ket{\varphi}\bra{\phi} \in B(\mathcal{H}) \longleftrightarrow 
\ket{\phi^{\ast},\varphi} \in \mathcal{H}\otimes \mathcal{H}\,.
\end{equation}
In particular, for an orthonormal basis of kets in $\mathcal{H}$,
$a=\bigl\{\ket{a_1},\ket{a_2}\dots,\ket{a_d}\bigr\}$, we have the conjugate
basis $a^*=\bigl\{\ket{a^*_1},\ket{a^*_2},\dots,\ket{a^*_d}\bigr\}$, and
jointly they are used in defining the two-qubit state
\begin{equation}
  \label{def-rhoa}
  \rho_a=\frac{1}{d}\sum_{j=1}^d\ket{a^*_ja^{\ }_j}\bra{a^*_ja^{\ }_j}\,,
\end{equation}
which has the $d$-fold eigenvalue $1/d$ and the $(d^2-d)$-fold eigenvalue zero.

We normalize the Hilbert-Schmidt inner product of two-qudit operators in
accordance with 
\begin{equation}
(A,B)=d\,\Tr{A^{\dagger}B}\,,  
\end{equation}
so that $(\rho_a,\rho_a)=1$ and $(\rho_a,\rho_b)=1/d$ for a pair of unbiased
bases. 
For the two-qudit states associated with two single-qudit bases, we then have
\begin{equation}
  \label{inner-ab}
  (\rho_a,\rho_b)=\frac{1}{d}\sum_{j,k=1}^d\bigl|\braket{a_j}{b_k}\bigr|^4
=1-\frac{d-1}{d}D_{ab}^2
\end{equation}
with the distance $D_{ab}$ of Eq.~(\ref{distance}), where the identity 
$\braket{a^*_ja^{\ }_j}{b^*_kb^{\ }_k}=\bigl|\braket{a_j}{b_k}\bigr|^2$
is used.
It follows that $D_{ab}$ can be expressed in terms of the Hilbert-Schmidt norm
of ${\rho_a-\rho_b}$,
\begin{equation}
  \label{Dab-HSnorm}
  D_{ab}=\sqrt{\frac{1}{2}\frac{d}{d-1}}\,|\!|\rho_a-\rho_b|\!|
\end{equation}
with $|\!|A|\!|=\sqrt{(A,A)}$.
This tells us something important: 
If $a\neq b$, then $\rho_a\neq\rho_b$, so that the
mapping $a\leftrightarrow\rho_a$ is one-to-one.

In passing, we note the following challenge. 
Clearly, not all two-qudit states with $\rho=d\rho^2$ correspond to a
single-qudit basis in the sense of Eq.~(\ref{def-rhoa}).
But which additional criteria identify the set of two-qudit states that do?

We are interested in finding the maximum value of the ASD 
between $k$ bases in dimension $d$. 
The numerical search begins with a randomly chosen initial set of bases, and
then modifies the bases in each iteration round such that $\avD$ is
systematically increased.

An infinitesimal variation of a ket in basis $a$ is given by
\begin{equation}
\delta\ket{a_j}=i \epsilon_a\ket{a_j},
\end{equation}
where $\epsilon_a$ is an infinitesimal hermitian operator acting on the basis
$a$. 
We have one such hermitian $\epsilon$ operator for each basis.
The resulting response of $\avD$ is 
\begin{equation}
  \delta\avD=\sum_{a=1}^k\tr{\epsilon_aG_a}\,,
\end{equation}  
where $\tr{\ }$ is a single-qudit trace and 
\begin{equation}
  G_a=\frac{8}{k(k-1)(d-1)}\,\mathrm{Im}\Biggl\{\sum_{b=1}^k\sum_{j,k=1}^d
      \bigl(\ket{a_j}\braket{a_j}{b_k}\bra{b_k}\bigr)^2\Biggr\}
\end{equation}
is the $a$th component of the gradient. 
If bases $a$ and $b$ are unbiased, there is no contribution to $G_a$ from
basis $b$ and, therefore, there is no gradient for a set of MUB.
But the converse is not true: We can have a vanishing gradient although the
bases are not pairwise unbiased.

When the gradient has nonzero components, we choose $\epsilon_a=\kappa G_a$
with a common $\kappa>0$ that specifies the step size.
This guarantees $\delta\avD>0$ if $\kappa$ is not too large, and maximization
along the line specified by the direction of the gradient can be done by
optimizing the value of $\kappa$.
The line optimization is a necessary ingredient if conjugate gradients are
used for accelerating the convergence; 
see Ref.~\cite{shewchuck94}, for instance.

The finite unitary change of basis $a$, 
$\ket{a_j}\to V_a\ket{a_j}$, is then accomplished by 
\begin{eqnarray}
  V_a&=&e^{i\epsilon_a}\nonumber\\
\mbox{or}\ 
  V_a&=&\frac{\mathbf{1}+i\epsilon_a/2}{\mathbf{1}-i\epsilon_a/2}\nonumber\\
\mbox{or}\ 
  V_a&=&(\mathbf{1}+i\epsilon_a)\prod_{n=0}^{\infty}
        \Bigl[\mathbf{1}+e^{i2\pi/3}\bigl(\epsilon_a^2\bigr)^{3^n}\Bigr]
\end{eqnarray}
or yet other ones, whichever of them is convenient to use.
All three $V_a$s equal $\mathbf{1}+i\epsilon_a$ 
to first order in $\epsilon_a$
and differ in the higher-order terms. 
Note that a high-precision evaluation of the infinite product in the third
version of $V_a$ requires very few terms.
This makes the third version a
viable alternative if the computation of the exponential in the first version
or of the inverse operator in the second version is time consuming or
imprecise.    

The iteration is terminated, when all components of the gradient vanish (in the
numerical sense specified by the machine precision).
We repeat this steepest-ascent search many times to ensure that we find the
global maximum. 
As Fig.~\ref{fig:histo} shows for $(d,k)=(6,4)$, the iteration gets stuck in
local maxima for about three attempts in ten and, see Table~\ref{tab:table1},
only four in ten trials are successful for $(d,k)=(6,7)$.

\section{Derivation of the two-parameter family}\label{app2}

The $d\times d$ matrix $U_{ab}$ composed of the transition amplitudes
$\braket{a_j}{b_k}$ of two orthonormal bases is unitary,
\begin{eqnarray}
  &&U_{ab}^{\ }=\left[
    \begin{array}{l}
      \bra{a_1}\\ \bra{a_2}\\ \vdots \\ \bra{a_d}
    \end{array}\right]
\Bigl[\ket{b_1},\ket{b_2},\ldots,\ket{b_d}\Bigr]=U_{ba}^{\dagger}\,,
\nonumber\\&& U_{ab}^{\ }U_{ba}^{\ }=\openone\,.
\end{eqnarray}
The columns and the rows of $U_{ab}$ are representations of the kets
$\ket{b_k}$ and the bras $\bra{a_j}$, respectively.
The unitary matrices associated with a set of bases 
have a composition law for consecutive basis changes:
$U_{ab}=U_{ac}U_{cb}$, $U_{aa}=\openone$.
In particular, $\sqrt{d}\,U_{ab}$ is a complex Hadamard matrix if the bases $a$
and $b$ are unbiased; see the paragraph containing Eq.~(\ref{Hada}).

Now, from the numerical search we know that one of the bases that maximize the
ASD between four bases in dimension six is unbiased with the other three
bases.
We identify this privileged basis as the canonical basis and refer to it as
the zeroth basis, and characterize the
set of four bases by the three $6\times6$ transition matrices
\begin{equation}
 M_1=U_{01}\,,\quad   M_2=U_{02}\,,\quad   M_3=U_{03}\,,
\end{equation}
so that the columns of $M_i$ are composed of the probability amplitudes of the
kets of the $i$th basis with respect to the privileged basis.

When multiplied by $\sqrt{6}$, the matrices $M_1$, $M_2$, and $M_3$ are
$6\times6$ Hadamard matrices, for which we use Karlsson's 
parameterization~\cite{karlsson10}.

His parameterization applies to $H_2$-reducible Hadamard matrices that can be
written in the form $H=X_L P_L N P_R X_R$, where the left and right $X$
matrices only contain phases on the diagonal, the $P$ matrices are permutation
matrices, and the central matrix has the form 
\begin{eqnarray}
N=\begin{bmatrix}
	F_2&T_1&T_2\\ 
  T_3&\frac{1}{2} T_3 A T_1&\frac{1}{2} T_3 B T_2\\ 
  T_4&\frac{1}{2} T_4 B T_1&\frac{1}{2} T_4 A T_2
\end{bmatrix},
\end{eqnarray}
where $F_2$ is the unnormalized two-dimensional Fourier matrix of
  Eqs.~(\ref{2x2a}) and the $2\times2$ $T_i$ matrices are 
those of Eq.~(\ref{2x2Zi}),
\begin{equation}
T_i=\begin{bmatrix}
	1&t_i\\ 1&-t_i
\end{bmatrix} \quad \textrm{with} \quad |t_i|=1\,,
\end{equation}
while
\begin{eqnarray}
A&=& F_2 \biggl(-\frac{1}{2} \openone + i \frac{\sqrt{3}}{2} \Lambda\biggr),
\nonumber \\ 
B&=& F_2 \biggl(-\frac{1}{2} \openone - i \frac{\sqrt{3}}{2} \Lambda\biggr)
\end{eqnarray}
with a unitary and hermitian $2\times2$ matrix $\Lambda$.
It turns out that our Hadamard matrices are indeed $H_2$-reducible since they
can be written as ${M_i=X_{L_i} P_{L_i} N_i P_{R_i} X_{R_i}}$ with the central
matrices given by
\begin{eqnarray}
N_1&=&\frac{1}{\sqrt{6}}
\begin{bmatrix}
	F_2&F_2&F_2\\
  F_2&\omega F_2&\omega^* F_2\\
  T&\omega^* T&\omega T
\end{bmatrix}\,, \nonumber\\
N_2&=&\frac{1}{\sqrt{6}}
\begin{bmatrix}
	F_2&F_2&F_2\\
  T&\omega T&\omega^* T\\
  T&\omega^* T&\omega T
\end{bmatrix}\,, \nonumber\\
N_3&=&\frac{1}{\sqrt{6}}
\begin{bmatrix}
	F_2&F_2&F_2\\
  T&\omega T&\omega^* T\\
  F_2&\omega^* F_2&\omega F_2
\end{bmatrix}\,;
\end{eqnarray}
see Eqs.~(\ref{M1}).

As in Eqs.~(\ref{2x2a}), we choose to express the matrix $T$ with factors of
$\omega=\exp(i2\pi/3)$,
\begin{equation}
T=\begin{bmatrix}
  1&\omega t^2\\
  1&-\omega t^2
\end{bmatrix},
\end{equation}
to exhibit the crucial dependence on the phase factor $t$.
The left permutation matrices are all equal, $P_{L_1}=P_{L_2}=P_{L_3}=P_L$.

Third, we notice that only the left dephasing and permutation matrices are
relevant for the distance. 
Indeed  the right dephasing matrices only add global phases to the basis
vectors while the right permutation only permute the basis vectors. 
In other words, two bases $B$ and $B P_R X_R$ are equivalent in terms of
distance. 
Therefore we can choose to conserve only the relevant structure for our bases,
that is, $M_i=X_{L_i} P_{L_i} N_i$. 

The fourth step is to use the fact that only relative dephasing and
permutations of the rows are relevant to the distance. 
Therefore we define new bases as
\begin{eqnarray}
M_1^{\ } &\widehat{=}& P_L^\dagger X_2^\dagger X_1^{\ } P_L^{\ } N_1^{\ }\,, 
\nonumber \\
M_2^{\ } &\widehat{=}& N_2^{\ }\,, \nonumber \\
M_3^{\ } &\widehat{=}& P_L^\dagger X_2^\dagger X_3^{\ } P_L^{\ } N_3^{\ }\,.
\end{eqnarray}
To simplify the notations, we again denote the two new diagonal matrices in
$P_L^\dagger X_2^\dagger X_1 P_L$ and $P_L^\dagger X_2^\dagger X_3 P_L$ by
$X_1$ and $X_3$, respectively. 
We further observe that
\begin{eqnarray}
X_1=\begin{bmatrix}
  A_1& \mathbf{0} & \mathbf{0}\\
  \mathbf{0} &A_2& \mathbf{0}\\
  \mathbf{0} & \mathbf{0} & A_1
\end{bmatrix}
\quad \textrm{and} \quad
X_3=\begin{bmatrix}
  B_1& \mathbf{0} & \mathbf{0}\\
  \mathbf{0} &B_2&\mathbf{0} \\
  \mathbf{0} & \mathbf{0} & B_3
\end{bmatrix}.\quad
\end{eqnarray}

Next we add a suitable global phase to $X_1$ and $X_3$. 
We multiply $X_1$ by $\exp(-i \textrm{Arg}(A_1[1,1]A_1[2,2]/2))$ and $X_3$ by
$\exp(-i \textrm{Arg}(B_1[1,1]B_1[2,2]/2))$ such that $A_1$ and $B_1$ take the
simple form 
\begin{eqnarray}
\begin{bmatrix}
  \exp(-i \phi)& 0\\
  0&\exp(i \phi)
\end{bmatrix},
\end{eqnarray}
for some phase $\phi$. 
We end up with the remarkable form
\begin{equation}
X_1=\begin{bmatrix}
  A_1& \mathbf{0} & \mathbf{0}\\
    \mathbf{0}&A_2&  \mathbf{0}\\
    \mathbf{0}& \mathbf{0} & A_1
\end{bmatrix}\quad\mbox{and}\quad
X_3=\begin{bmatrix}
  A_1^*& \mathbf{0} & \mathbf{0}\\
    \mathbf{0}&\omega^*A_1^*&  \mathbf{0}\\
    \mathbf{0}& \mathbf{0} & B_3
\end{bmatrix}
\end{equation}
where [cf.\ Eqs.~(\ref{2x2a})]
\begin{equation}
A_1=\begin{bmatrix}
  x^*&0\\
  0&x
\end{bmatrix}.
\end{equation}
So far, we have found that
\begin{eqnarray}
A_3 &=& A_1\,, \nonumber \\
B_1 &=& A_1^*\,, \nonumber \\
B_2 &=& \omega^* A_1^*\,,
\end{eqnarray}
and it only remains to find the structure behind the two $2\times2$ dephasing
matrices $A_2$ and $B_3$. 

To do so, we now consider the products $U_{ij}=M_i^\dagger M_j^{\ }$. 
We obtain
\begin{eqnarray}
M_1^\dagger M_2^{\ }=
\begin{bmatrix}
  a_1 & a_2 & a_3\\
  a_3 & a_1 & a_2\\
  a_2 & a_3 & a_1
\end{bmatrix}
\quad \textrm{with} \quad
\begin{bmatrix}
     a_1\\
     a_2\\
     a_3
\end{bmatrix}=F_3
\begin{bmatrix}
  F_2^{\ } A_1^* F_2^{\ }\\
  F_2^{\ } A_2^* T\\
  T^\dagger A_3^* T
\end{bmatrix} \nonumber \\
\end{eqnarray}
and $F_3$ is the standard (unnormalized) 3-dimensional Fourier matrix
\begin{eqnarray}
F_3=\begin{bmatrix}
      1& 1 & 1\\
      1 & \omega & \omega^*\\
      1 & \omega^* & \omega
    \end{bmatrix}.
\end{eqnarray}
Similarly we have
\begin{eqnarray}
M_2^\dagger M_3^{\ }=
    \begin{bmatrix}
      b_1 & b_2 & b_3\\
      b_3 & b_1 & b_2\\
      b_2 & b_3 & b_1
\end{bmatrix}
\quad \textrm{with} \quad
\begin{bmatrix}
     b_1\\
     b_2\\
     b_3\\
\end{bmatrix}=F_3
\begin{bmatrix}
     F_2 B_1 F_2\\
     T^\dagger B_2 T\\
     T^\dagger B_3 F_2
\end{bmatrix} \nonumber \\
\end{eqnarray}
and
\begin{eqnarray}
M_3^\dagger M_1^{\ }=
\begin{bmatrix}
      c_1 & c_2 & c_3\\
      c_3 & c_1 & c_2\\
      c_2 & c_3 & c_1
\end{bmatrix}
\quad \textrm{with} \quad
\begin{bmatrix}
     c_1\\
     c_2\\
     c_3\\
\end{bmatrix}=F_3
\begin{bmatrix}
     F_2 Y_1 F_2\\
     T^\dagger Y_2 F_2\\
     F_2 Y_3 T
\end{bmatrix} \nonumber \\
\end{eqnarray}
where
\begin{eqnarray}
Y=X_3^* X^{\ }_1=
\begin{bmatrix}
      Y_1 & \mathbf{0} & \mathbf{0}\\
      \mathbf{0} & Y_2 & \mathbf{0}\\
      \mathbf{0} & \mathbf{0} & Y_3
\end{bmatrix}
=\begin{bmatrix}
      A_1^2 & \mathbf{0} & \mathbf{0}\\
      \mathbf{0} & \omega A_1 A_2 & \mathbf{0}\\
      \mathbf{0} & \mathbf{0} & B_3^*A_1^{\ }
\end{bmatrix}. \nonumber \\
\end{eqnarray}

The seventh step is to look once more at the numerics. 
With respect to the product $M_1^\dagger M_2^{\ }$, we see that
\begin{eqnarray}
a_2=\omega^* Z a_3 Z\,.
\end{eqnarray}
Thus we are lead to define the matrix equation
\begin{eqnarray}
E_1 \widehat{=} a_2-\omega^* Z a_3 Z=0\,.
\end{eqnarray}
This only represents a system of three equations since $E_1[1,1]=E_1[2,2]$. 
In the same manner, we have for $M_2^\dagger M_3^{\ }$
\begin{eqnarray}
E_2 \widehat{=} b_1-\omega^* Z b_3 Z=0\,,
\end{eqnarray}
and $E_2[1,1]=E_2[2,2]$ so that, here too, only three equations are relevant. 
Finally, for $M_3^\dagger M_1^{\ }$, we obtain
\begin{eqnarray}
E_3 \widehat{=} c_1+Z c_2 Z=0
\end{eqnarray}
and, owing to $(\omega^*-1)E_3[1,2]= t (1- \omega) E_3[2,1]$, again only
three equations are relevant.  
We should mention here that there are other interesting identities within the
products $M_i^\dagger M_j^{\ }$, such as 
$b_2=[a_1+a_1^\dagger+Z(a_1-a_1^\dagger) Z]/2$,
but they are much more complicated to handle and will not be necessary to
achieve our parameterization. 

The eighth step is to solve the above nine equations. 
We obtain
\begin{eqnarray}
E_1[1,1]&:& \; \tr{A_1}=\tr{A_3}\,, \nonumber \\
E_1[1,2]&:& \; A_1-2 \omega^* t^{*2} A_2+\omega^* t^{*2} A_3= r \openone\,, 
\nonumber \\
E_1[2,1]&:& \; \omega^* t^{*2}A_1-2 \omega^* t^{*2} A_2+A_3= r' \openone\,.\quad
\end{eqnarray}
From the numerics, we know that $r=r'$ and thus $A_1=A_3$, 
which we already found by looking at the dephasing matrix $X_1$. 
Note also that the expression of the complex number $r$ is not required. 
Furthermore we find
\begin{eqnarray}
E_2[1,1]&:& \; \tr{B_1}=\omega \tr{B_2}\,, \nonumber \\
E_2[1,2]&:& \; \omega^* t^{*2}B_1+ \omega B_2-2 \omega t^{*2}A_3= s\openone\,, 
\nonumber \\
E_2[2,1]&:& \;  B_1+ t^{*2} B_2 -2 \omega t^{*2} B_3= s' \openone\,.
\end{eqnarray}
From the numerics, we know that $s=s'(=r)$ and thus $B_1=\omega B_2$, which we
already obtained by looking at the dephasing matrix $X_3$. 
The next three equations are much more interesting. 
Indeed we have
\begin{eqnarray}
E_3[1,1]&:& \; 2 \tr{Y_1}-\omega^* \tr{Y_2}-\omega \tr{Y_3}=0\,, \nonumber \\
E_3[2,2]&:& \; 2 \tr{Y_1}-\omega t^{*2} \tr{Y_2}-\omega^* t^2 \tr{Y_3}=0\,, 
\nonumber \\
E_3[1,2]&:& \;  t^{*2} Y_2 -Y_3= u \openone\,.
\end{eqnarray}
From the numerics, we know that $u=0$ and the last equation reduces to
\begin{equation}
Y_3=t^{*2}Y_2.
\end{equation}
Since $Y_2=\omega A_1 A_2$ and $Y_3=B^*_3A_1$, 
the above equation directly translates into
\begin{equation}
B_3=\omega^* t^{2} A_2^*\,.
\end{equation}
This last relation can be inserted in $E_3[1,1]$ and $E_3[2,2]$, which become
identical and can be written as
\begin{eqnarray}\label{soon}
2 \tr{Y_1}-(\omega^* + \omega t^{*2}) \tr{Y_2}=0\,.
\end{eqnarray}
This equation will soon become Eq.~(\ref{fame}).

A last hint from the numerics is needed. We actually notice that
\begin{equation}
Y_1 Y_2 Y_3=- \openone\,.
\end{equation}
As $Y_3=t^{*2}Y_2$, we arrive at $t^{*2}Y_1 Y_2^2=-\openone$ so that $\omega
t^{*} A_1^2 A_2=\pm i U$, where $U^2=\openone$, that is, $U=\openone$ or
$U=Z$ since it has to be diagonal. 
With the help of the numerics, we conclude that
\begin{equation}
A_2= i \omega^* t Z A_1^{*2}
\end{equation}
and consequently
\begin{equation}
B_3= -i t Z A_1^{2}\,.
\end{equation}
The final parametrization of the dephasing matrices is therefore given by
\begin{eqnarray}
X_1&=&\begin{bmatrix}
      A_1&  \mathbf{0}& \mathbf{0}\\
      \mathbf{0} & i\omega^* t Z A_1^{*2}& \mathbf{0}\\
      \mathbf{0} &\mathbf{0}  & A_1
\end{bmatrix}\,, \nonumber \\
X_3&=&\begin{bmatrix}
      A_1^*& \mathbf{0} & \mathbf{0}\\
      \mathbf{0} &\omega^* A_1^*& \mathbf{0}\\
      \mathbf{0} & \mathbf{0} & -i t Z A_1^{2}
\end{bmatrix} \,,
\end{eqnarray}
which are ingredients in Eqs.~(\ref{M1}).

Let us finally come back to Eq.~(\ref{soon}). 
We can now substitute $Y_1=A_1^2$ and $Y_2=(i \omega^* t Z A_1^{*2})(\omega
A_1)=i t Z A_1^*$ in Eq.~(\ref{soon}) and, upon defining
$x=\exp(i \theta_x)$ and $t=\exp(i \theta_t)$, we arrive at 
\begin{eqnarray}
\cos(\theta_t-2\pi/3)=-\frac{\cos(2 \theta_x)}{\sin(\theta_x)}\,,
\end{eqnarray}
which is Eq.~(\ref{fame}).

\newcommand{\Title}[1]{\textit{#1}, }  

\end{document}